\documentclass[times, 5p, number, pdftex]{elsarticle}
\usepackage{amsmath} 
\usepackage{amssymb} 
\usepackage{amsfonts}
\usepackage{dsfont}
\usepackage{graphicx} 
\usepackage{color}
\usepackage[breaklinks]{hyperref}
\usepackage[latin2]{inputenc}
\usepackage{hyperref}
\usepackage{textcomp}
\usepackage{subfig}
\usepackage{floatrow}

\begin{document}

\begin{frontmatter}
\title{A Copula Approach on the Dynamics of Statistical Dependencies in the US Stock Market}

\author[dui,bos]{Michael C. M\"unnix\corref{cor1}}
\ead{michael@muennix.com}
\author[dui]{Rudi Sch\"afer}
\address[dui]{Fakult\"at f\"ur Physik, Universit\"at Duisburg-Essen, 47048 Duisburg, Germany}
\address[bos]{Center of Polymer Studies, Department of Physics, Boston University, USA}
\cortext[cor1]{Corresponding author. Tel.: +49 203 379 4727; Fax: +49 203 379 4732.}

\begin{abstract}
We analyze the statistical dependency structure of the S\&P 500 constituents in the 4-year period from 2007 to 2010 using intraday data from the New York Stock Exchange's TAQ database. With a copula-based approach, we find that the statistical dependencies are very strong in the tails of the marginal distributions. This \emph{tail dependence} is higher than in a bivariate Gaussian distribution, which is implied in the calculation of many correlation coefficients. We compare the tail dependence to the market's average correlation level as a commonly used quantity and disclose an nearly linear relation.
\end{abstract}

\end{frontmatter}
\section{Introduction}
The measurement of statistical dependence is often broken down to the calculation of a correlation coefficient, such as the Pearson coefficient \cite{pearson00} or the Spearman coefficient \cite{spearman87}. Correlation coefficients are widely used in various disciplines of science. It is also often included in financial modeling, e.g., in the Capital Assets Pricing Model (CAPM) \cite{sharpe64} or Noh's model \cite{noh00}.

The usage of the correlation coefficient, however, suggests a  the linear statistical dependence and that the observables are nearly normal distributed. Due to the central limit theorem, this might be justified in some cases, but often the statistical dependence is much more complex. In these cases, the statistical dependence cannot be represented by a single number. 
The joint probability distribution, of course, holds all information of the statistical dependence. Certainly, the joint probability distribution also contains the individual marginal probability distributions. These can have different shapes depending on the underlying process. The statistical dependence of different systems usually cannot be directly compared with this approach.

Copulae, first introduced by Sklar in 1959 \cite{sklar59, sklar73}, permit a separation between the pure statistical dependence and the marginal probability distributions. This allows to compare the statistical dependence of diverse systems. 

The usage of copulae is well established in statistics and finance; There are many classes of analytical copula functions that meet various properties \cite{b_nelseon98}. Several studies of financial markets are devoted to developing suitable copluae or fitting existing ones to empirical data \cite{fernandez08, chavezdemoulin05, rosenberg06} or are based on a small subset of assets \cite{malevergne03}. In this study, we chose a different approach. We perform a large-scale empirical study to disclose the structure of the average pairwise copula of the US stock returns. As the copula does not depend on the shape of the return distribution, we are able to average over the copula of different stock pairs although their marginal distributions' shape may differ, i.e., exhibits stronger or weaker tails. In particular, we study the intraday stock market returns of the 428 continuous S\&P 500 constituents in 2007--2010 based on intraday data from the New York Stock Exchange's TAQ database.

\section{Copulae}
The basic concept is simple: Let $a$ and $b$ be two random variables with probability densities $f_{a}(x)$ and $f_{b}(x)$ and cumulative distributions $F_{a}(x)$ and $F_{b}(x)$, with
\begin{align}
\int\limits_{-\infty}^{+\infty}f_{a}(x) &= 1\ ,\\
F_{a}(x) &= \int\limits_{-\infty}^{x}f_{a}(x')\ dx'\ ,
\end{align}
and analogously for $b$. Further, let $f_{a,b}(x,y)$ be the joint probability density and $F_{a,b}(x,y)$ be the joint cumulative distribution. The inverse cumulative distribution function $F^{-1}$ is the called the quantile function. For example, $F^{-1}_{a}(0.05)$ represents the value which 5\% of all random samples are smaller or equal to. This evidently gives,
\begin{equation}
F_{a}\left(F^{-1}_{a}(\alpha)\right)=\alpha \ .
\end{equation}
$F^{-1}(\alpha)$ is also called the $\alpha$-quantile. The copula $\mathrm{Cop}_{a,b}(u,v)$ is defined as the cumulative joint distribution of quantiles, 
\begin{align}
\mathrm{Cop}_{a,b}(u,v) = F_{a,b}\left(F_{a}^{-1}(u),F_{b}^{-1}(v)\right) \ .  
\end{align}
The copula density $\mathrm{cop}_{a,b}(u,v)$ is consequently defined by
\begin{align}
\mathrm{cop}_{a,b}(u,v) = \frac{\partial^{2}}{\partial u \partial v}\mathrm{Cop}_{a,b}(u,v) \ .
\end{align}
As the quantile functions $F^{-1}$ are scale free, the copula does not depend on the underlying marginal distributions. It only contains the pure statistical dependence. Thus, by obtaining the appropriate copula of a system, one can simply interchange the marginal distributions without any changes in the copula. This is very useful if the marginal distributions change for some reason, but the statistical dependence remains the same. We can rebuild the joint cumulative distribution from the copula and the individual distributions by
\begin{equation}
F_{a,b}(x,y)=\mathrm{Cop}_{a,b}\left(F_{a}(x),F_{b}(y)\right).
\end{equation}

\begin{figure}[tp]
\centering
\subfloat[]{

\begin{minipage}{1\textwidth}
\centering
$\vcenter{\hbox{\includegraphics[width=0.88\textwidth]{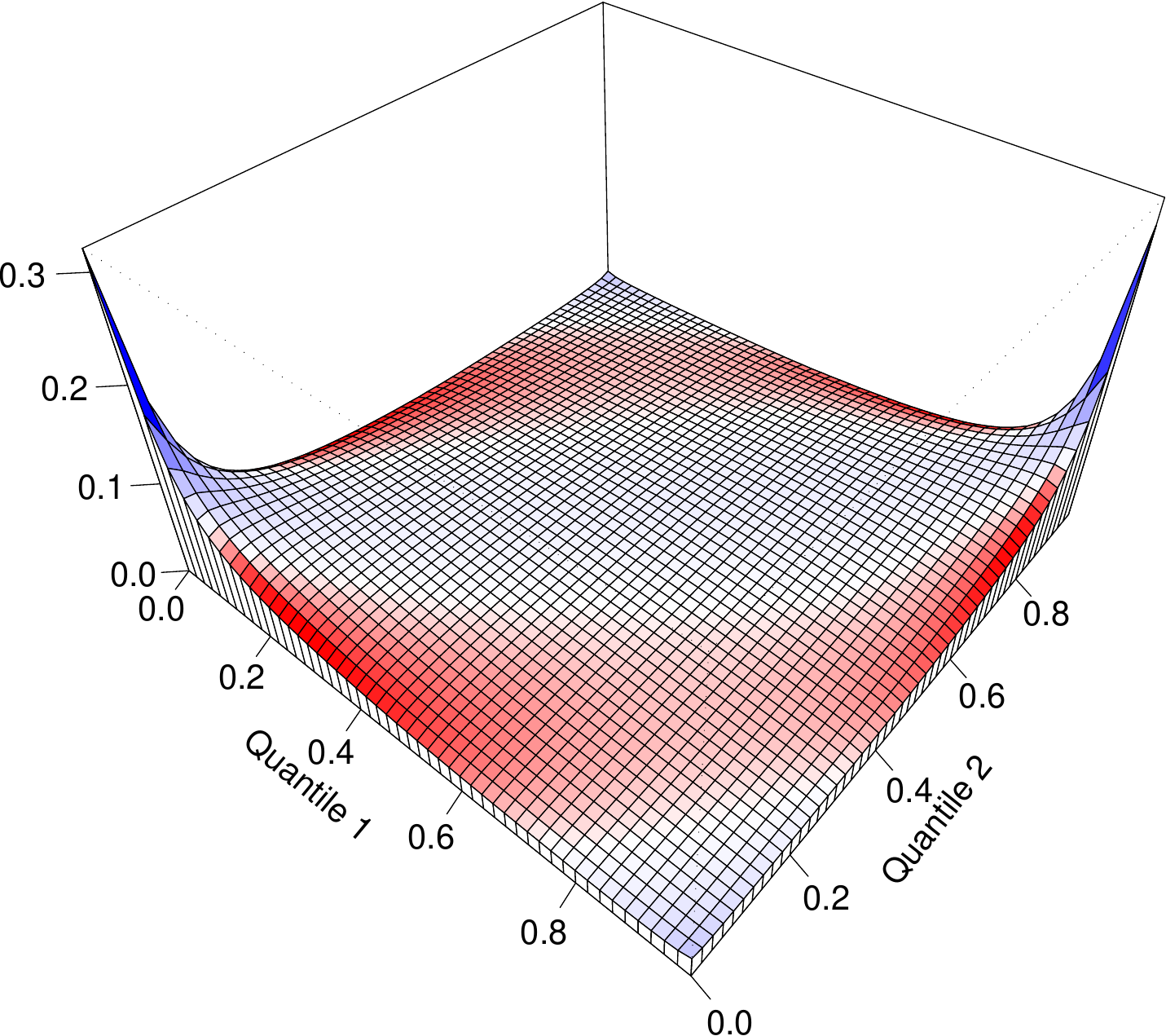}}}$
\hspace{1mm}
$\vcenter{\hbox{\includegraphics[width=0.09\textwidth]{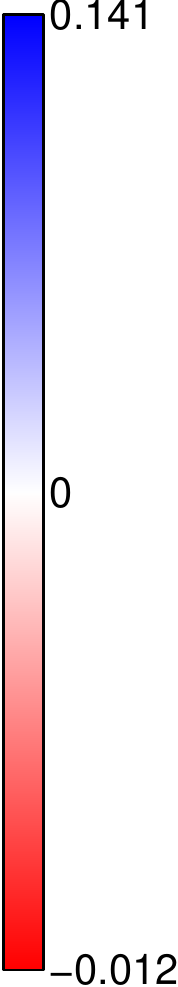}}}$

\end{minipage}

\label{fig:copulaaveragegauss}
}
\\
\subfloat[]{\includegraphics[width=1\textwidth]{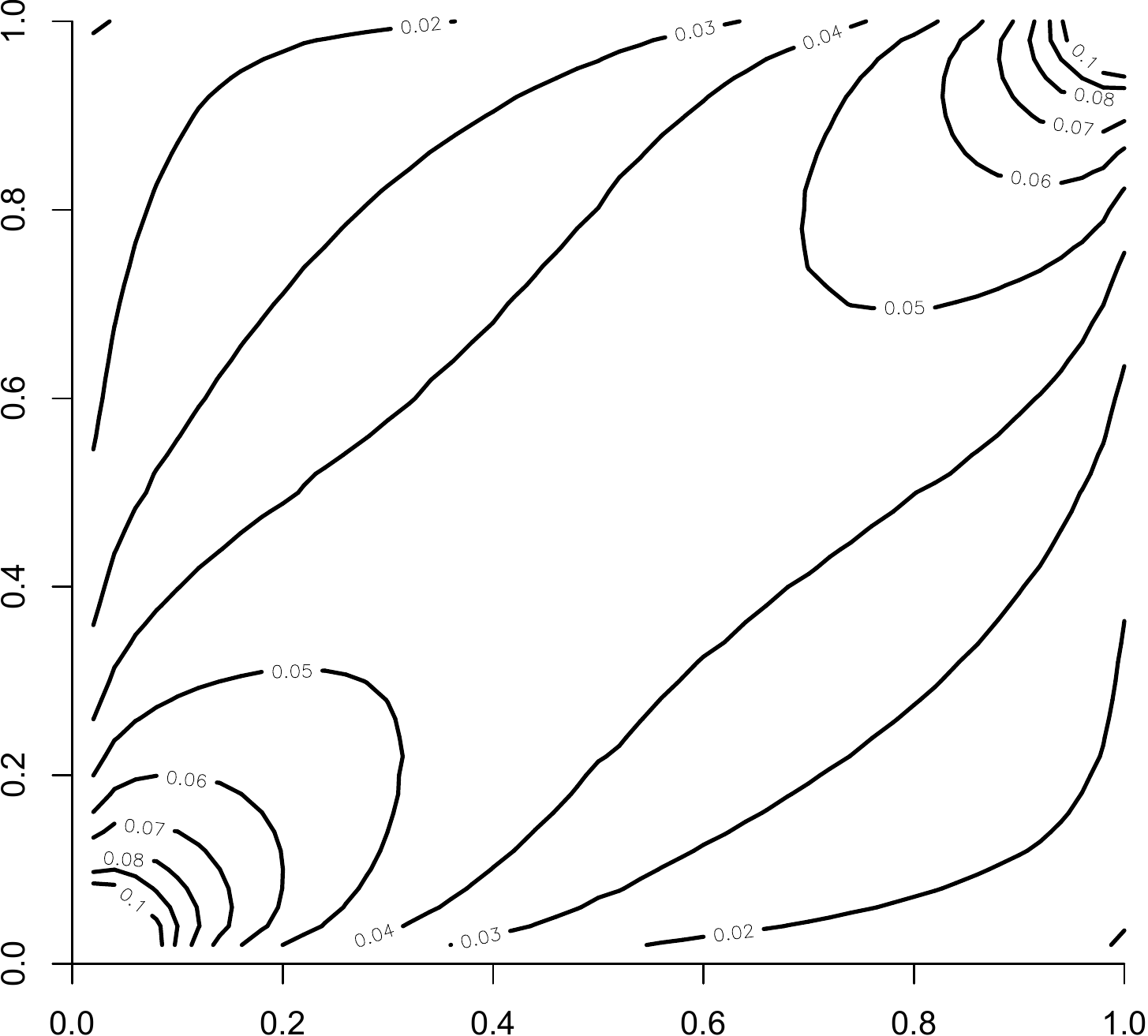}}
\caption{Average pairwise copula of the S\&P 500 stock returns in 2007--2010. The z-axis in (a) and the isolines in (b) are in permille. The color shading in (a) illustrates the difference to the Gaussian copula (positive values mean that Gaussian copula is less dense).}
\label{fig:copulaaverage}
\end{figure}

\section{Average copula}
To calculate the cumulative copula from empirical data of two return time series $r_{1}$ and $r_{2}$, we use 
\begin{align}
\mathrm{Cop}_{r_{1},r_{1}}(u,v)= \frac{1}{T}\sum\limits_{t=1}^{T} \mathrm{1_{U}}(r_{1}(t)) \times \mathrm{1_{V}}(r_{2}(t))\ ,
\end{align}
where $T$ is the length of the time series. $\mathrm{1_{U}}$ and $\mathrm{1_{V}}$ are indicator functions relating to the sets
\begin{align}
U &= \left\{x\ |\ x \le F_{1}^{-1}(u) \right\}\ , \\  
V &= \left\{y\ |\ y \le F_{2}^{-1}(v) \right\}\ .
\end{align}
The quantile function $F^{-1}$ on empirical data is given by
\begin{align}
F^{-1}_{1}(u)=
\begin{cases}
\mathrm{inf}\left\{ x\ |\  F_{1}(x) \ge u \right\}\quad &0<u\le1 \\
\mathrm{sup}\left\{ x\ |\  F_{1}(x) = u \right\}\quad &u=0
\end{cases}\ ,
\end{align}
and analogously for $r_{2}$. We define $F_{1}(x)$ empirically as the percentage of the portion that is smaller or equal to $x$ compared to the total amount of values. When calculating the empirical copula density, it is useful to first define a resolution of the 2D grid, e.g. $m=50$. On this $m\times m$ grid, we can calculate the copula by
\begin{align}
\mathrm{cop}_{r_{1},r_{1}}\left(\frac{i}{m},\frac{j}{m}\right)= \frac{1}{T}\sum\limits_{t=1}^{T} \mathrm{1_{\bar U_{i}}}(r_{1}(t)) \times \mathrm{1_{\bar V_{j}}}(r_{2}(t)) \quad i,j \in {1\dots m}\
\end{align}
with 
\begin{align}
\bar U_{i} &= \left\{x\ \Big|\ F_{1}^{-1}\left(\frac{i-1}{m}\right) < x \le F_{1}^{-1}\left(\frac{i}{m}\right) \right\}\ , \\  
\bar V_{j} &= \left\{y\ \Big|\  F_{2}^{-1}\left(\frac{j-1}{m}\right) < y \le F_{2}^{-1}\left(\frac{j}{m}\right)\right\}\ .
\end{align}
Of course, an accurate estimation of the copula density requires a large amount of data points. Thus, we estimate the average copula using intraday data. We start with the calculation of 30-minute arithmetic returns, because market microstructure distortions dominate at smaller return intervals \cite{epps79,muennix09b,muennix10a}. We expand our analysis further by calculating 1-hour, 2-hour and 4-hour returns.
\begin{figure}[tb]
\centering
\begin{minipage}{1\textwidth}
\centering
$\vcenter{\hbox{\includegraphics[width=0.88\textwidth]{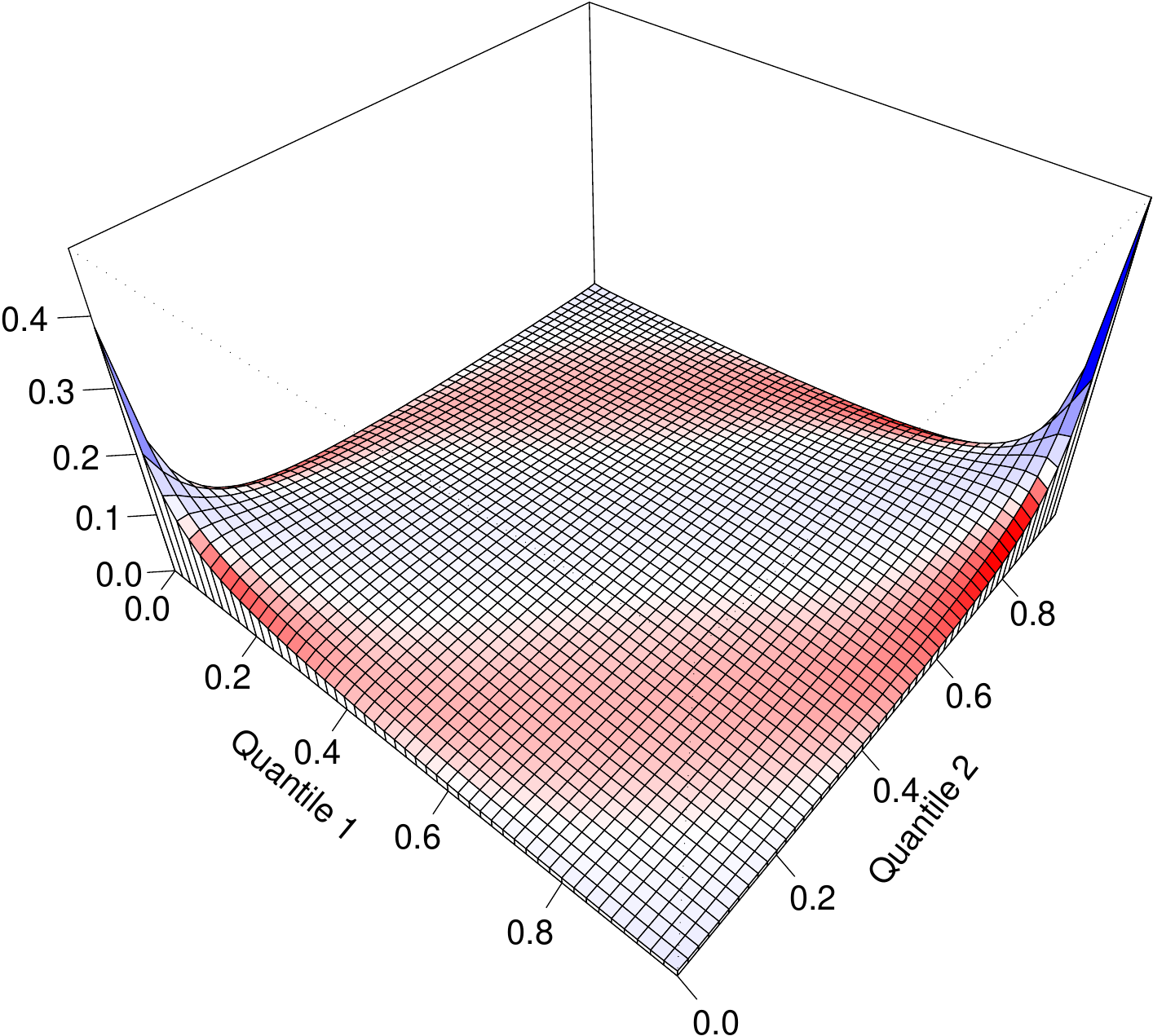}}}$
\hspace{1mm}
$\vcenter{\hbox{\includegraphics[width=0.09\textwidth]{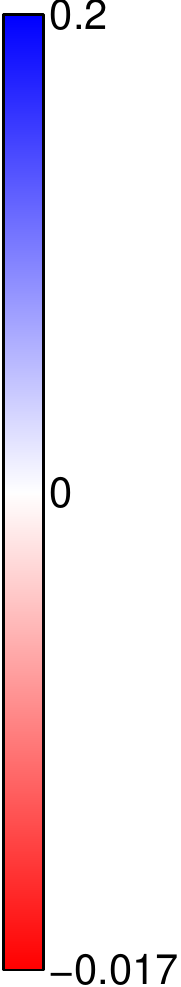}}}$
\end{minipage}
\caption{Average pairwise copula of the S\&P 500 stock returns in during the crisis period from 2008/10/15 to 2009/4/1.}
\label{img:crisiscopula}
\end{figure}
We obtain a very similar copula for all return intervals. This is very surprising, because it is well-known that the shape of the marginal return distribution changes towards small return intervals -- the tails of the distributions become stronger \cite{gopikrishnan99, plerou99b}. However, apparently this does not change the statistical dependence.
The results are shown in figure \ref{fig:copulaaverage}, exemplarily for 1-hour returns. The copula has high density in the outer quantiles. This corresponds to a higher correlation in the tails of the return distribution than in it's center. This is often referred to as \emph{tail dependence} \cite{frahm05,schmidt06,heffernan00}. Our results indicate that on average, the upper tail dependence is stronger than the lower tail dependence. For comparison, the average difference to the Gaussion copula (which is implied by many correlation coefficients) is illustrated in figure \ref{fig:copulaaveragegauss}. The (standard normal) Gaussian Copula is given by
\begin{align}
\mathrm{Cop}_{c}(u,v)&=F_{c}(F^{-1}(u),F^{-1}(v))\ ,\\
\mathrm{cop}_{c}(u,v)&=\frac{f_{c}(F^{-1}(u),F^{-1}(v))}{f(F^{-1}(u)) f(F^{-1}(v))} \ . 
\end{align}
Here, $f_{c}$ and $F_{c}$ refer to the bivariate standard normal probability density and cumulative distribution with correlation $c$. $f$ is the univariate standard normal probability density, while $F^{-1}$ is the corresponding quantile function. To calculate the average difference $d$, we have to calculate the Gaussian copula based on all coefficients of the correlation matrix $\mathbf{C}$, based on $K=428$ stocks and subtract it from the empirical copula,
\begin{align}
d(u,v)=\frac{\sum\limits_{i=1}^{K}\sum\limits_{j=i+1}^{K}\left( \mathrm{cop}_{i,j}(u,v) - \mathrm{cop}_{C_{i,j}}(u,v) \right)}{K(K-1)/2} \ .
\end{align}
This gives us information about how erroneous the dependence is estimated if implying a Gaussian copula.
The empirical copula exhibits a stronger dependence than the Gaussian copula. The probability of correlated extreme events is underestimated. The lower tail dependence is stronger than the upper tail dependence. There is general a trend that this behavior is more pronounced towards large return intervals. This might be caused by a more severe reaction on bad news than on good news. We will discuss this in more detail in the next section. Another feature of the empirical copula is the relatively high density in the (0,1) and (1,0) corners, indicating the presence of anti-correlated extreme events.

Figure \ref{img:crisiscopula} illustrates the copula during the market meltdown between 2008 and 2009. Surprisingly it exhibits a stronger positive tail dependence than negative tail dependence. However, the main observation is much higher here. The assumption of the Gaussian Copula would have been a dramatic mistake during this period. The Gaussian copula is even being discussed for having a main impact of the financial crisis \cite{lee09}.

\begin{figure*}[tb]
\begin{center}
\includegraphics[width=0.8\textwidth]{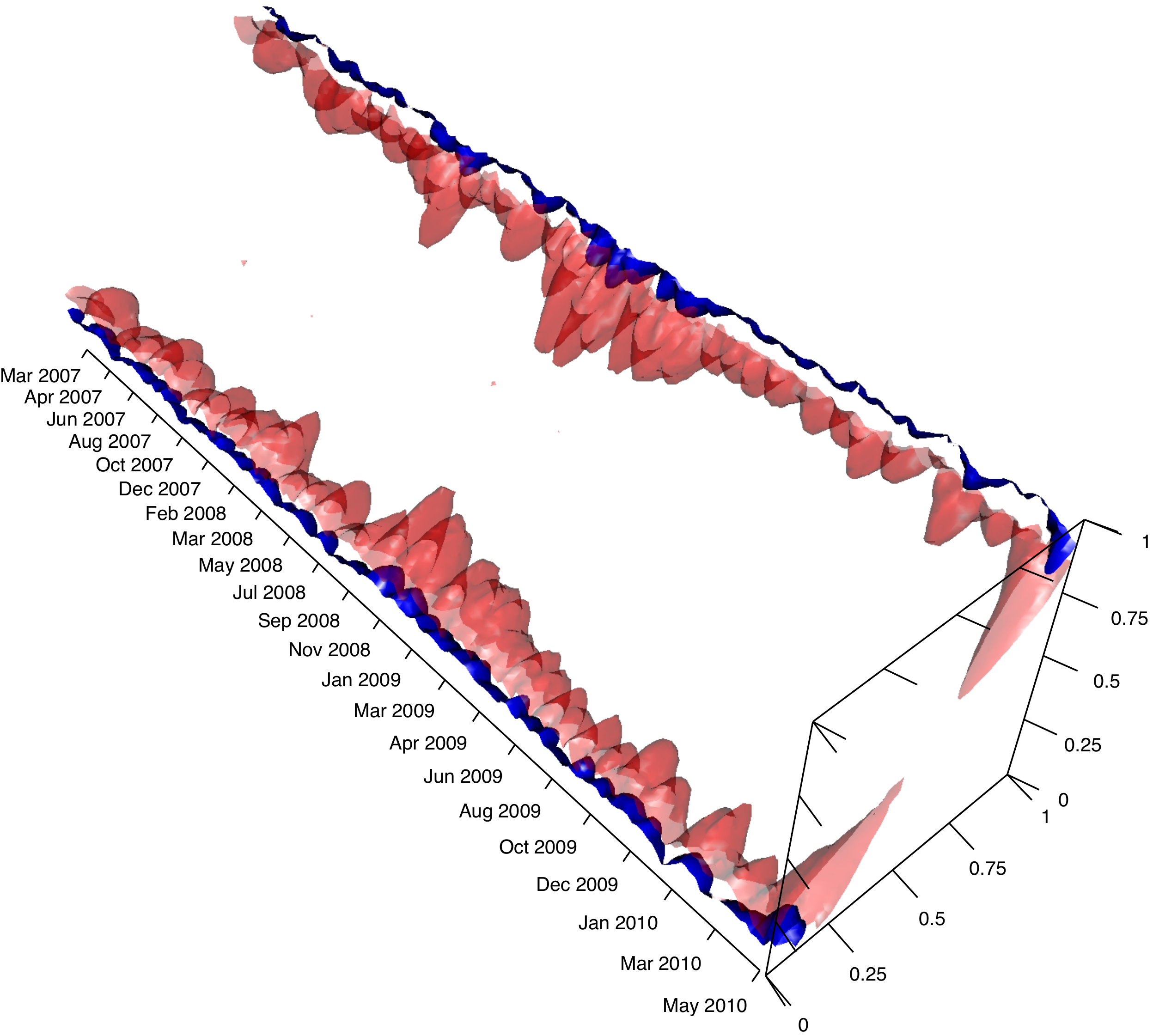}
\caption{Evolution of the S\&P 500 stocks' average pairwise copula density. The isosurfaces correspond to a probability of 0.1\textperthousand~(blue) and 0.05 \textperthousand~(red). The density in the tails is very high.}
\label{fig:copuladynamic}
\end{center}
\end{figure*}

\section{Dynamics of the copula}
It is evident that statistical dependencies of financial assets change in time. For example, this can be caused by microeconomic influences, changing political factors or herding effects. Several studies address this issue with the concept of correlation coefficients \cite{tastan06, rosenow03, kullmann02, drozdz01}. Here, we approach this matter with an empirical study of the changes in the average pairwise copula. We calculate the average copula within 2-week periods within the 2007-2010 period based on 1-hour returns. Results are shown in figure \ref{fig:copuladynamic}. To illustrate the structural changes of the copula, we plot the isosurfaces in the tail regions. We discover that the tail dependence is stronger during financial crashes, such as from Oct 2008 to Feb 2010. But the fluctuations of the tail dependence are very large. It reflects the current market's situation in a sensible manner.

Often financial crashes are accompanied by overall very large correlation coefficients. This raises the question if there is some dependence between the market's average correlation level and the tail dependence. To obtain an insight into this question we compare the average correlation coefficient of the whole market in each 2-week period to the tail dependence. As correlation coefficients are still widely used, this maps a correlation coefficient to one of the most important features of the copula.

\begin{figure*}[p]
\centering
\subfloat[$\Delta t = 30\mathrm{min}, \alpha=0.02$]{\includegraphics[width=0.35\textwidth]{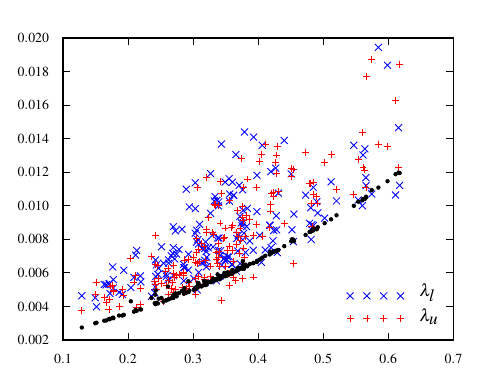}}
\subfloat[$\Delta t = 30\mathrm{min}, \alpha=0.04$]{\includegraphics[width=0.35\textwidth]{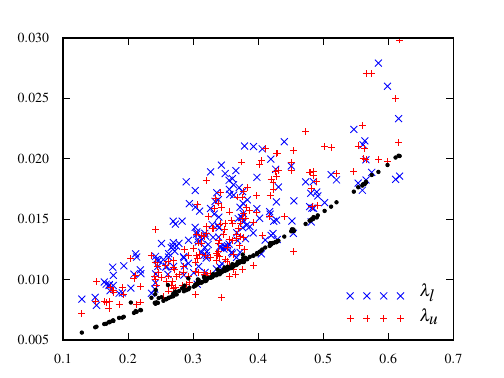}}\\
\subfloat[$\Delta t = 30\mathrm{min}, \alpha=0.1$]{\includegraphics[width=0.35\textwidth]{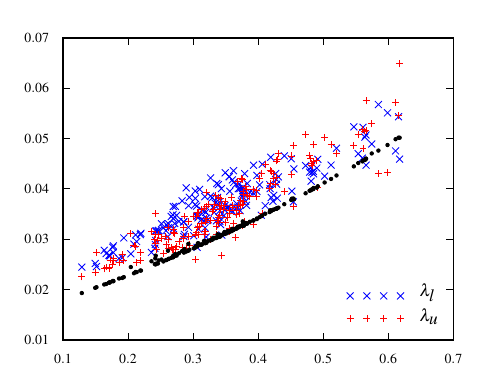}}
\subfloat[$\Delta t = 30\mathrm{min}, \alpha=0.25$]{\includegraphics[width=0.35\textwidth]{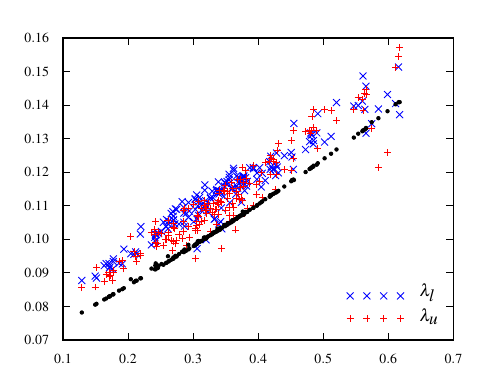}}\\
\subfloat[$\Delta t = 60\mathrm{min}, \alpha=0.02$]{\includegraphics[width=0.35\textwidth]{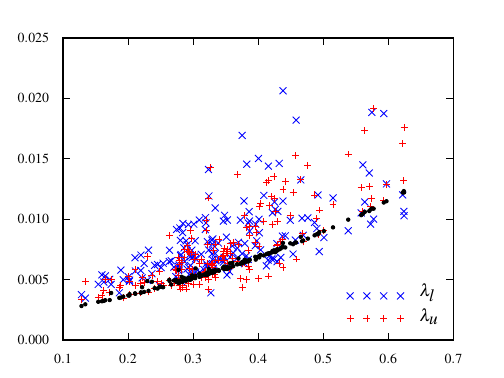}}
\subfloat[$\Delta t = 60\mathrm{min}, \alpha=0.04$]{\includegraphics[width=0.35\textwidth]{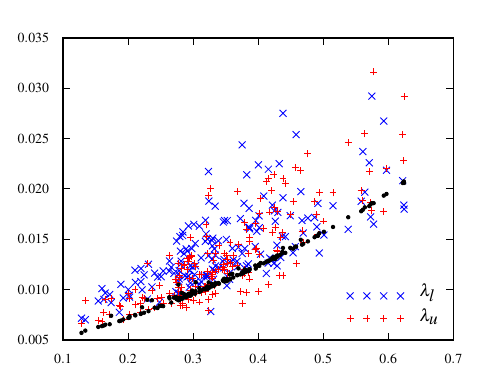}}\\
\subfloat[$\Delta t = 60\mathrm{min}, \alpha=0.1$]{\includegraphics[width=0.35\textwidth]{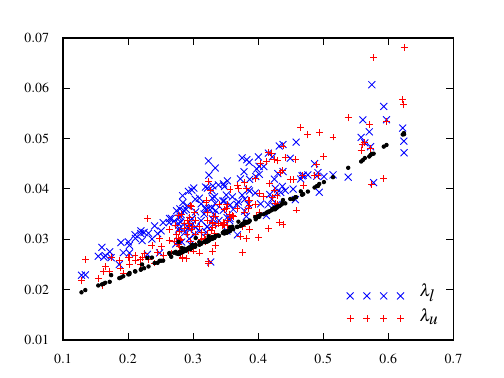}}
\subfloat[$\Delta t = 60\mathrm{min}, \alpha=0.25$]{\includegraphics[width=0.35\textwidth]{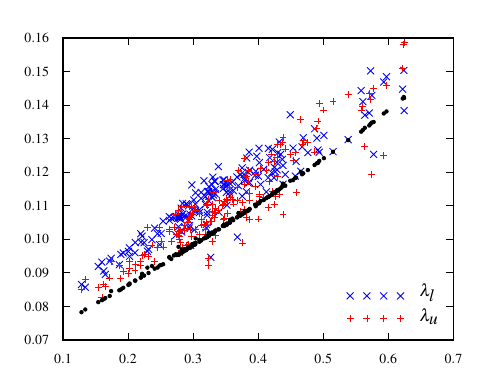}}
\caption{Relation between tail dependence and average correlation level for different quantiles $\alpha$ and return intervals $\Delta t$.}
\label{fig:corr-vs-c}
\end{figure*}

\begin{figure*}[p]
\ContinuedFloat 
\centering
\subfloat[$\Delta t = 120\mathrm{min}, \alpha=0.02$]{\includegraphics[width=0.35\textwidth]{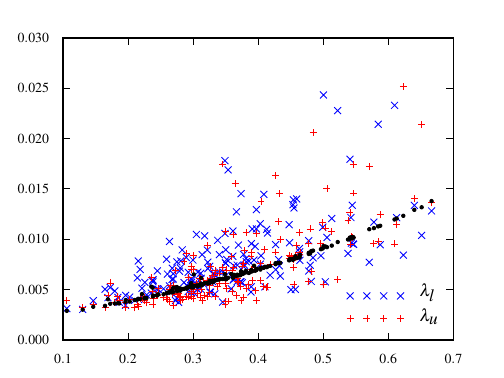}}
\subfloat[$\Delta t = 120\mathrm{min}, \alpha=0.04$]{\includegraphics[width=0.35\textwidth]{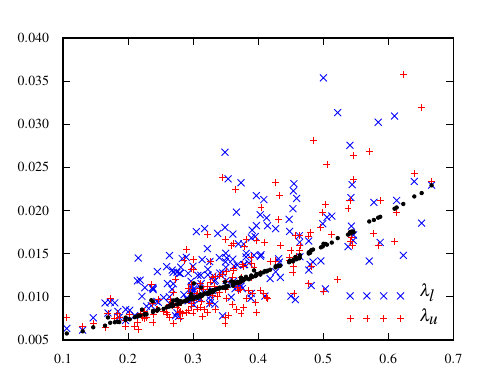}}\\
\subfloat[$\Delta t = 120\mathrm{min}, \alpha=0.1$]{\includegraphics[width=0.35\textwidth]{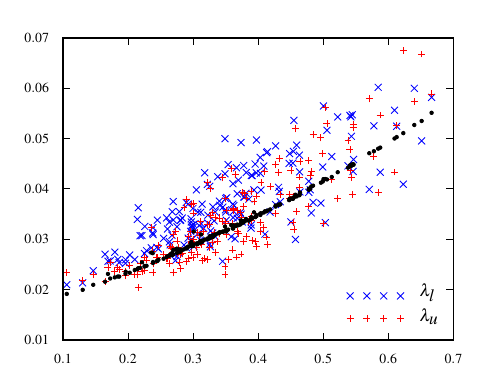}}
\subfloat[$\Delta t = 120\mathrm{min}, \alpha=0.25$]{\includegraphics[width=0.35\textwidth]{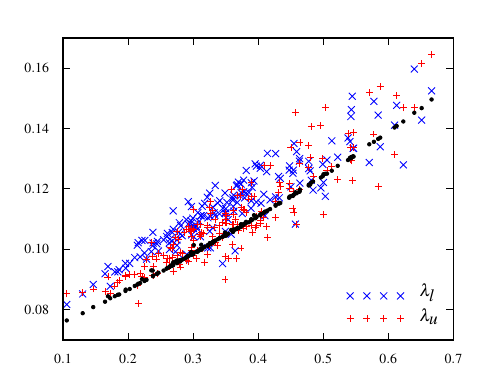}}\\
\subfloat[$\Delta t = 240\mathrm{min}, \alpha=0.02$]{\includegraphics[width=0.35\textwidth]{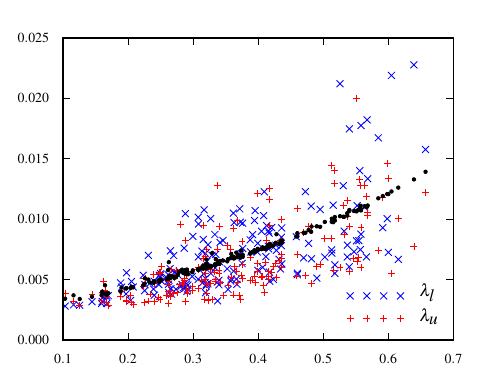}}
\subfloat[$\Delta t = 240\mathrm{min}, \alpha=0.04$]{\includegraphics[width=0.35\textwidth]{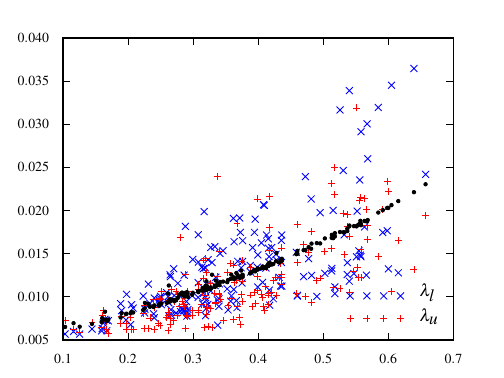}}\\
\subfloat[$\Delta t = 240\mathrm{min}, \alpha=0.1$]{\includegraphics[width=0.35\textwidth]{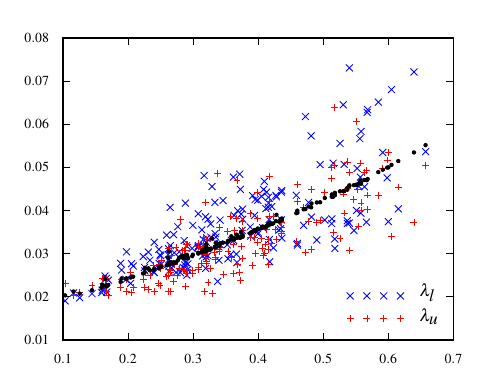}}
\subfloat[$\Delta t = 240\mathrm{min}, \alpha=0.25$]{\includegraphics[width=0.35\textwidth]{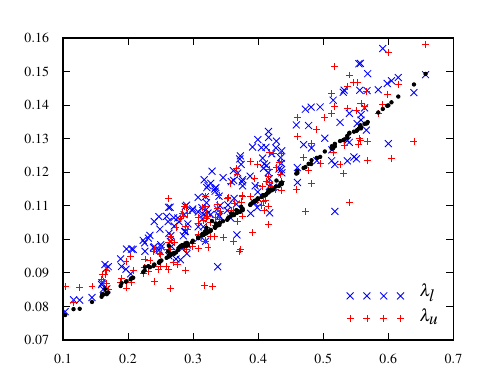}}
\caption{(continued)}
\label{fig:corr-vs-c}
\end{figure*}

To quantify this tail dependence, we calculate the probability of two returns to be simultaneously above or below a certain quantile $\alpha$. This very simple form of a upper and lower tail dependence coefficient is given by
\begin{align}
\lambda_{l}(\alpha)&=\mathrm{Cop}(\alpha,\alpha)\ , \\
\lambda_{u}(\alpha)&=1-\mathrm{Cop}(1-\alpha,1-\alpha)\ .
\end{align}
More advanced tail dependences are, e.g., discussed in Ref. \cite{heffernan00}. However, as we only examine the difference between the empirical copula and the Gaussian copula, we restrict ourselves to this measure. We perform the analysis for return intervals from 30 minutes to two hours.
Results are shown in figure \ref{fig:corr-vs-c}. We find a very strong relation of the tail dependence and the average correlation coefficient. For comparison we build the average tail dependence coefficients $\lambda_{l}$ and $\lambda_{u}$ of the Gaussian copula, given by
\begin{equation}
\lambda_{l}=\lambda_{u}=\mathrm{Cop}_{c}(\alpha,\alpha) \ .
\end{equation}
To calculate the average Gaussian tail dependence, for each 2-week period, we calculate the tail dependence of the Gaussian copula based on the correlation matrix' entries $C_{i,j}$ of this period,
\begin{align}
\left\langle\lambda^{\mathrm(Gauss)}_{l} \right\rangle = \left\langle\lambda^{\mathrm(Gauss)}_{u} \right\rangle
= \frac{\sum\limits_{i=1}^{K}\sum\limits_{j=i+1}^{K}\left(\mathrm{Cop}_{C_{i,j}}(\alpha,\alpha)\right)}{K(K-1)/2} \ .
\end{align}
This gives the opportunity to compare how the tail dependence is overall misjudged, if using correlation coefficients or the Gaussian copula.

The relation between the market's average correlation level and the tail dependence appears to be almost linear.
This is similar to the Gaussian copula except that the tail dependence is more pronounced.
For small return intervals, such as $\Delta t$ = 30min and 60min, the tail dependence has a tendency to be stronger than in the Gaussian case. For small quantiles, such as $\alpha$ = 2\% and 4\%, there are many cases where this linear relation does not hold. There are many outliers that feature a much stronger tail dependence than in the Gaussian case.
On larger return intervals, the tail dependence becomes more and more similar to the Gaussian case, which is consistent with studies of the marginal distributions \cite{gopikrishnan99}. Here, the lower tail dependence is significantly higher than the upper tail dependence, as discussed in the previous section. 
This underlines the unsuitability of the Gaussian copula for the estimation of \emph{correlated} extreme events. This is a key ingredient to the estimation of financial risk \cite{b_bouchaud00, schaefer07, rosenberg06, chavezdemoulin05}.

\section{Conclusion}
In a large scale empirical study of the S\&P 500 stock's copula, we disclosed important features of the dependence structure.
This gives the opportunity to isolate the statistical dependence structure from features of the probability distributions, such as heavy tails.
In general, the overall average pairwise copula of the 4-year feature stronger tails than the Gaussian copula. Extreme events are much more correlated than assumed by a linear correlation. Moreover, empirical copula indicates the presence of anti-correlated extreme events. Despite the large differences between the Gaussian marginal distribution and the distribution of high frequency returns, the dependency structure is quite similar.  
In a more detailed study, where we calculated the time-dependent empirical copula in the resolution of 2-weeks we showed that the Gaussian copula, in particular, systematically underestimates the negative tail dependence: The market reacts sensible to large negative returns resulting in a collective downward motion.
The evolution of the copula in the 4-year period discloses a strong relation between the market's average correlation level and the tail dependence. For return intervals of 4 hours and in the center region of the distribution, the Gaussian copula describes the situation fairy well. But when using smaller return intervals or estimating the tail regions, the fluctuations in the correlation-tail-dependence relation become very strong. 

\section*{Acknowledgements}
We thank O. Grothe for fruitful discussions.
M.C.M. acknowledges financial support from the Fulbright program and from Studienstiftung des deutschen Volkes.

\section*{References}
\bibliographystyle{elsarticle-num}
\bibliography{Manuscript_arxiv2.bbl}

\end{document}